\documentclass[%
 reprint,
 amsmath,amssymb,
 aps,
]{revtex4-1}

\usepackage{float} 
\usepackage{graphicx}
\graphicspath{{figures/}}
\usepackage{caption}
\usepackage{subcaption}
\usepackage{dcolumn}
\usepackage{bm}
\usepackage{multirow}
\usepackage{xfrac}
\usepackage[shortlabels]{enumitem}

\begin{document}


\title{Defining Standard Strategies for Quantum Benchmarks}

\author{Mirko Amico}
\email{mamico@ibm.com}
\author{Helena Zhang}
\author{Petar Jurcevic}
\author{Lev S. Bishop}
\author{Paul Nation}
\author{Andrew Wack}
\author{David C. McKay}
\email{dcmckay@us.ibm.com}
\affiliation{IBM Quantum, Yorktown Heights, NY 10598 USA}

\date{\today}

\begin{abstract}
As quantum computers grow in size and scope, a question of great importance is how best to benchmark performance. Here we define a set of characteristics that any benchmark should follow---randomized, well-defined, holistic, device independent---and make a distinction between benchmarks and diagnostics. We use Quantum Volume~(QV)~\cite{cross2019validating} as an example case for clear rules in benchmarking, illustrating the implications for using different success statistics, as in Ref.~\cite{baldwin:2022}. We discuss the issue of benchmark optimizations, detail when those optimizations are appropriate, and how they should be reported. Reporting the use of quantum error mitigation techniques is especially critical for interpreting benchmarking results, as their ability to yield highly accurate observables comes with exponential overhead, which is often omitted in performance evaluations. Finally, we use application-oriented and mirror benchmarking techniques to demonstrate some of the highlighted optimization principles, and introduce a scalable mirror quantum volume benchmark. We elucidate the importance of simple optimizations for improving benchmarking results, and note that such omissions can make a critical difference in comparisons. For example, when running mirror randomized benchmarking, we observe a reduction in error per qubit from 2\% to 1\% on a 26-qubit circuit with the inclusion of dynamic decoupling.
\end{abstract}

\maketitle


\section{\label{sec:intro}Introduction}
In the rapidly advancing field of quantum computing, it is essential to have methods in place to accurately measure the performance of quantum hardware, comparing devices across different technologies and gauging generational system improvements. This is squarely in the domain of benchmarks; a set of standardized routines used to evaluate device performance. However, there is no universal agreement on what constitutes an appropriate benchmark, what dimensions to measure against (speed, quality, and/or scale), and given the ubiquitous nature of noise in quantum devices, what methodologies are appropriate when reducing noise through various optimization techniques. Some such methodologies could constitute an unfair advantage when utilized to improve benchmarking, and their uneven application can result in unjust comparisons. Furthermore, benchmark results often come with hidden execution overheads and limits on scalability that must be examined critically.

In order to discuss benchmarks, we must first define what a comprises a valid benchmark. While there is no singular answer to this question, commonly accepted benchmarks, in general, share a specific feature set.  Namely they are composed of randomized circuit instances, follow well-defined execution procedures, are holistic (i.e. pertaining to  performance on as many attributes as possible) in nature, and may be executed on any gate-based quantum computing architecture. We give a few examples of benchmarks that satisfy these rules, like QV and CLOPS, and discuss the effects of using a different set of rules for the same benchmark as done in Ref.~\cite{baldwin:2022} for the success statistics of QV. Following the definition of a benchmark, we propose that there is a differentiation between what are true benchmarks and what are commonly termed benchmarks, but are in fact \textit{diagnostics}. In contrast to benchmarks that are designed to capture the overall performance of a device in an average sense, diagnostic methods are highly sensitive to only certain sources of error or specific device components. Almost all algorithmic circuits are diagnostic, and most accurately predict the success of a given circuit of the same structure. For this reason, there is utility in the rise of application-oriented circuit libraries, such as the ones contained in Refs. \cite{lubinski2021application, mesman2021qpack, finvzgar2022quark, tomesh2022supermarq, zhang2023characterizing, kurlej2022benchmarking, lubinski2023optimization, kordzanganeh2022benchmarking, mundada2022experimental,li:2022}. These aim to collect various algorithms with varying quantum circuit structures in an attempt to capture the performance of quantum hardware. Given that they average over a wide set of circuits, they may rise to the level of true benchmarks when executed in aggregate, depending on the interpretation of the outcomes. We will discuss the pros and cons of these proposals, and highlight the importance of being cautious when judging overall performance based on a small subset of applications. Furthermore, we will emphasize the role of mirror circuits introduced in Ref.~\cite{proctor2022scalable}, and generalized in Ref.~\cite{proctor2022measuring}, as a convenient method for circumventing some of the inherent limitations in application-based protocols.

Next, we consider what constitute appropriate optimizations of a benchmark. Quantum hardware performance necessarily needs to focus on the trinity of scale, quality, and speed \cite{wack2021quality}. Often when one refers to benchmarking methods, attention is focused on only one of these attributes, namely quality. However, all benchmarks must look at the trade-offs between these three attributes, as demonstrated in Ref.~\cite{lubinski2023optimization}, when evaluating the speed and quality aspects of performance. We emphasize a distinction between optimization techniques based on their resource requirement. On one hand, there is tremendous interest in emerging error mitigation techniques that trade off an exponential amount of speed for quality \cite{cai2022quantum,takagi2022fundamental} for their potential usage in reaching quantum advantage for some application. On the other hand, there are also a number of common error suppression techniques, such as dynamic decoupling~\cite{viola1999dynamical}, that can be implemented with near-constant algorithmic resource overhead. These techniques can also be used to optimize benchmarks, and, in fact, should always be included to the best of one's ability to implement them. Similarly, there are a class of pre-execution optimization routines, such as optimal gate decomposition and efficient layout selection techniques~\cite{nation2022suppressing}, that can yield important performance improvement with limited classical overhead. Disclosing the optimization techniques used will then be necessary for a fair comparison of benchmarks across devices and over time. Finally, there are optimization techniques that have no rational role in benchmarking, for example those that utilize the ideal output of the circuit to improve the experimentally obtained fidelity. We'll show an example of one of such technique, where result bitstrings are filtered based on their frequency of occurrence, in the last section of the manuscript. In all cases, it behooves the operators of such techniques to be clear and transparent about the optimizations utilized and the speed overhead they entail.

In the experimental component of this manuscript, we utilize the suite of applications proposed in Ref.~\cite{lubinski2021application}, and mirror circuits in Refs.~\cite{proctor2022scalable, proctor2022measuring}, to show the effect that error suppression and mitigation techniques have on the reported values. Even for constant-overhead techniques, such as dynamic decoupling \cite{viola1999dynamical, pokharel2018demonstration}, layout optimization, and measurement mitigation, we can see large improvements in both application and mirror benchmarking circuits. Using ever more sophisticated and resource intensive error mitigation techniques, it is possible to substantially improve the quality of results even further. Finally, we show an example of how utilizing the output of the circuit one can gain artificial advantage for application-oriented benchmarking circuits.\\

The article is structured in the following way: \S~\ref{sec:sqs} reviews the definition of a benchmark in the context of a protocol to measure the key attributes of scale, quality and speed. Here we define a set of postulates that every benchmark should follow, and differentiate from protocols which are merely diagnostic. We overview some of the common benchmarks, and demonstrate a mirror quantum volume benchmark. \S~\ref{sec:opt_bench} looks at the different ways in which benchmarks can be optimized, laying some rules for how optimizations that trade off the scale, quality and speed need to be reported. Finally, in \S~\ref{sect:examples} we run some application-circuits from the QED-C benchmark suite and mirror circuits to illustrate the principles discussed.

\section{\label{sec:sqs} Defining Benchmarks}

With the growth of the size and complexity of quantum computers, it becomes increasingly important to have methods in place to determine progress and compare different devices. Furthermore, those methods need to consider three key attributes: scale, quality, and speed. Scale refers to the size (e.g. in number of qubits) of the problems that can be solved. Importantly, the number of qubits effectively used for solving a problem may be lower than the physical qubits involved in the computation. For example, this is the case when certain error mitigation techniques are used \cite{koczor2021exponential}. Quality is a measure of how faithfully a quantum computer executes a specific task, with deviations from ideal behaviour signalling the presence of one or more sources of error. Finally, speed represents the temporal performance of a system, measured either via the number of operations executed per unit of time, or by computing the run-time of a full quantum algorithm. Together these attributes measure the progress of a device, with an eye to the relevant aspects which make an error-corrected device; a system requiring a large number of qubits to encode logical qubits, with low enough error to be below the code threshold and with the ability to complete code checks quickly \cite{babbush2021focus}. They also dictate the relative utility of a device, i.e., the algorithmic power. \\

Given these attributes, what are the methodologies for determining progress---these are benchmarks. However, it's important that benchmarks satisfy certain criteria in order for them to be faithful and fair measures of these attributes across widely different technology platforms. Here we propose an overarching set of rules for benchmarks:
\begin{itemize}
    \item \textbf{Random}: Tests should have a randomized component (e.g. the circuits, input and/or outputs), and the final metric aggregated over this randomization used to measure an average result.
    \item \textbf{Well-defined}: Benchmarks should have a clear set of rules for defining and running the protocol, so that there is no ambiguity, and others can reproduce the procedure \footnote{If the source of randomness in a benchmark is classical, then it is possible to seed a pseudo-random number generator, and fully recover the original experiment.  However, this is not the case when the source of entropy is quantum in nature, e.g. as done for CLOPS, and reproducing results make sense only with ensemble averaging.}.
    \item \textbf{Holistic}: The benchmarking results should be indicative of performance over a large set of the device attributes in as few metrics as possible, i.e., scale should be implicit in the benchmark.
    \item \textbf{Platform independent}: The protocol should not be tailored to a particular gate-set and should be independent of the types of connectivity and native gates of the platform, so long as it conforms to a universal gate set and the circuit model of quantum computing.
\end{itemize}
These protocols should have reasonable scalability as well, although at present this isn't an overriding concern. We also note that such scalability can be easily addressed by mirror circuits, which we discuss in \S~\ref{sec:mirror}. Note that in addition to reducing ambiguity, clear rules will ensure that there are no shortcuts that can circumvent the spirit of the benchmark, for example, compiling to a null circuit. However, these rules do not have to explicitly address other kinds of optimization that we will later discuss in \S~\ref{sec:opt_bench}, such as error mitigation strategies.   \\


Another point that we want to emphasize in this discussion is that there are still a large number of circuits that can be executed on hardware (e.g. from the application suites~\cite{lubinski2021application, mesman2021qpack, finvzgar2022quark, tomesh2022supermarq, zhang2023characterizing, kurlej2022benchmarking, lubinski2023optimization, kordzanganeh2022benchmarking, mundada2022experimental,li:2022}), and the output of those circuits are extremely useful in understanding device performance. In particular, the randomization postulate for a benchmark means that the benchmarks may not have high predictability with respect to specific algorithmic circuits and, for example, this can lead to situations where the observed performance differs from what is predicted. While application circuits and their proxies test the performance of the device on specific applications, which is useful for characterizing the device and predicting performance for similar circuits, it is important to be cautious when judging overall performance based on a small subset of structured circuits. This is especially true given that no application currently holds significant real-world value and so strong claims cannot be made for any particular circuit to be more relevant. The choice among them only translates to sampling circuits with different structures which may highlight particular strengths or weaknesses that don't generalize to all tasks. This has been recently discussed in Refs.~\cite{wang2022sok, quetschlich2022mqt}, where distinctions between the layer of abstractions of benchmarking are considered.

In light of these challenges, it is important to make a clear differentiation between methods that we refer to as benchmarks and methods that we refer to as diagnostics. We propose the following definitions:
\begin{itemize}
 \item \textbf{Benchmark}: As defined in our criteria above.
 \item \textbf{Diagnostic}: A protocol (circuits and output success measure) that is highly sensitive to certain types of errors, e.g., Hellinger fidelity of GHZ states.
\end{itemize}

While benchmarks are designed to compare across technologies and device iterations, diagnostic methods can be used to have a clear characterization of performance in a particular setting. The result of diagnostic methods should be highly predictive for similarly structured problems. This is the case for most application-inspired methods, which can give a precise indication of expected performance on specific tasks (or similarly structured tasks). However, because of their specificity, diagnostic methods are not good standards. Even when collecting together a suite of diagnostic methods, it is hard to determine whether these will cover all aspects of a device's performance and so benchmark suites must be carefully constructed. Also, in the context of getting the maximum performance out of quantum hardware in a specific application, it may be desirable to use compilation and mitigation techniques, thus making diagnostic methods good candidates for including such techniques in their execution. This is in contrast with benchmarking methods, where we are interested in characterizing the average performance we can expect to achieve when executing a randomly chosen task. These issues will be addressed in the next section.

\subsection{Example Benchmarks}

Here we discuss some relevant quality and speed benchmarks that adhere to the postulates mentioned above, while noting that this is far from an exhaustive list. We discuss four examples: quantum volume~\cite{cross2019validating}, CLOPS~\cite{wack2021quality}, mirror RB/circuits~\cite{proctor2022measuring} and application-suites. For QV we specifically discuss this in the context of the benchmarking postulate of having clear rules. 

\subsubsection{Quantum volume}
A widely used quality benchmark is quantum volume (QV) \cite{cross2019validating}, a single-number metric which represents the useful space-time volume of circuits that can be successfully executed on a given system. Denoting the quantum volume as $V_Q$, its expression takes the following form:

\begin{equation}
    \log_2 V_Q = \text{argmax}_m \; \text{min}\left( m, d(m) \right)
\end{equation}

\noindent
for \textbf{Haar random} circuits of width $m$ and achievable depth $d(m)$. Where the achievable depth is the highest value of $d$ such that the heavy-output probability (HOP) $h_U$ is greater than $2/3$, where the heavy-outputs are those bit-strings that have probabilities above the median value of the distribution obtained by classically simulating the random circuit $U$. Since benchmarks must abide by the clear rules postulate, all QV experiments must abide by rules as laid out in Table.~\ref{tab:qv_rules}. Among those rules is a test on the statistics of the test passing, where $\sigma$ was defined strictly in Ref.~\cite{cross2019validating} by the conditions
\begin{align}
    n_c&\ge100\\
    \frac{n_h-2\sqrt{n_h(n_s-n_h/n_c)}}{n_c n_s}&>\frac{2}{3}  ,
\end{align}
where there were $n_h$ observed heavy outputs from a total of $n_c n_s$ trials using $n_s$ shots on each of $n_c$ random circuits. This condition was motivated by the observation that for any trial of a random protocol there are risks of both false positives (meeting the benchmark when the system performance does not justify passing) and false negatives (not meeting the benchmark despite sufficient system performance). Any finite-duration protocol must have a smooth transition between consistently failing at low performance and consistently succeeding at high performance. We will always be striving to achieve the next barely-attainable benchmark and therefore false-negatives are especially pernicious because it is easy to game the benchmark by making many attempts and waiting to get a lucky false-positive. This can happen even if you do not set out to game in this way, since you can easily convince yourself that fine-tuning some experiment parameters is the secret to the success rather than the repeated attempts implied by that fine tuning. With this framework in place the QV success criterion has the following properties:
\begin{enumerate}
    \item Bound the false positive rate from above at a small `$2\sigma$' rate $\epsilon\simeq0.02275$;
    \item Ensure that for passing parameters one can push the false-negative rate arbitrarily low at the cost of increased protocol time due to increased $n_c$;
    \item Force a minimum effort on running the protocol to avoid rapid attempts to brute-force obtaining a false positive result.
\end{enumerate}

\begin{figure}
    \includegraphics[width=\columnwidth]{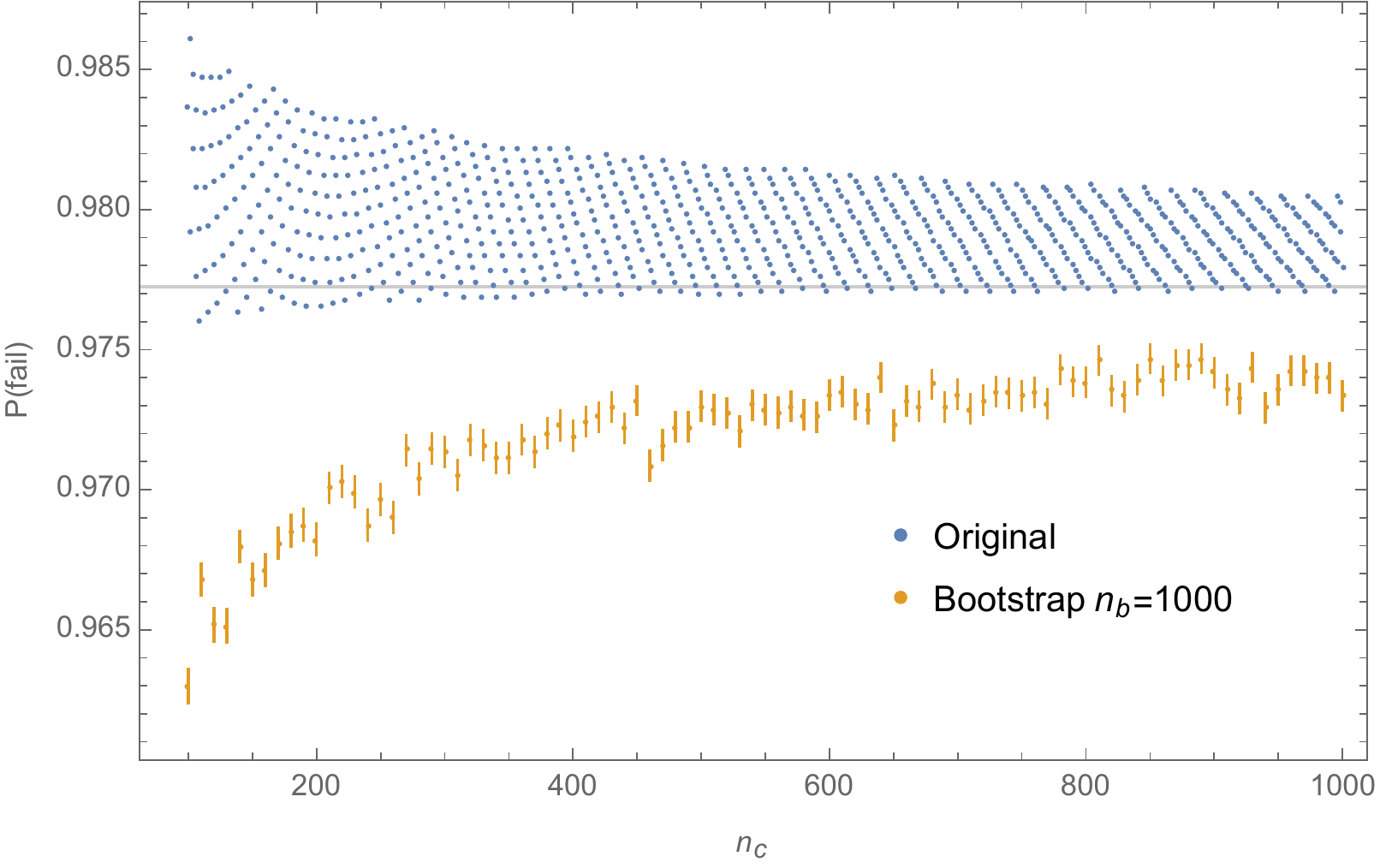}
    \caption{Failure rate just below threshold for the original~\cite{cross2019validating} and bootstrapped~\cite{baldwin:2022} protocols, as a function of $n_c$ the number of circuits sampled. The original protocol keeps the false-positive rate capped at the scale of the target $\epsilon$ (denoted by horizontal gridline) whereas the bootstrap protocol can exceed by a factor of up to $\sim1.6$.}
    \label{fig:negativeratevsnc}
\end{figure}

\begin{figure}
    \includegraphics[width=\columnwidth]{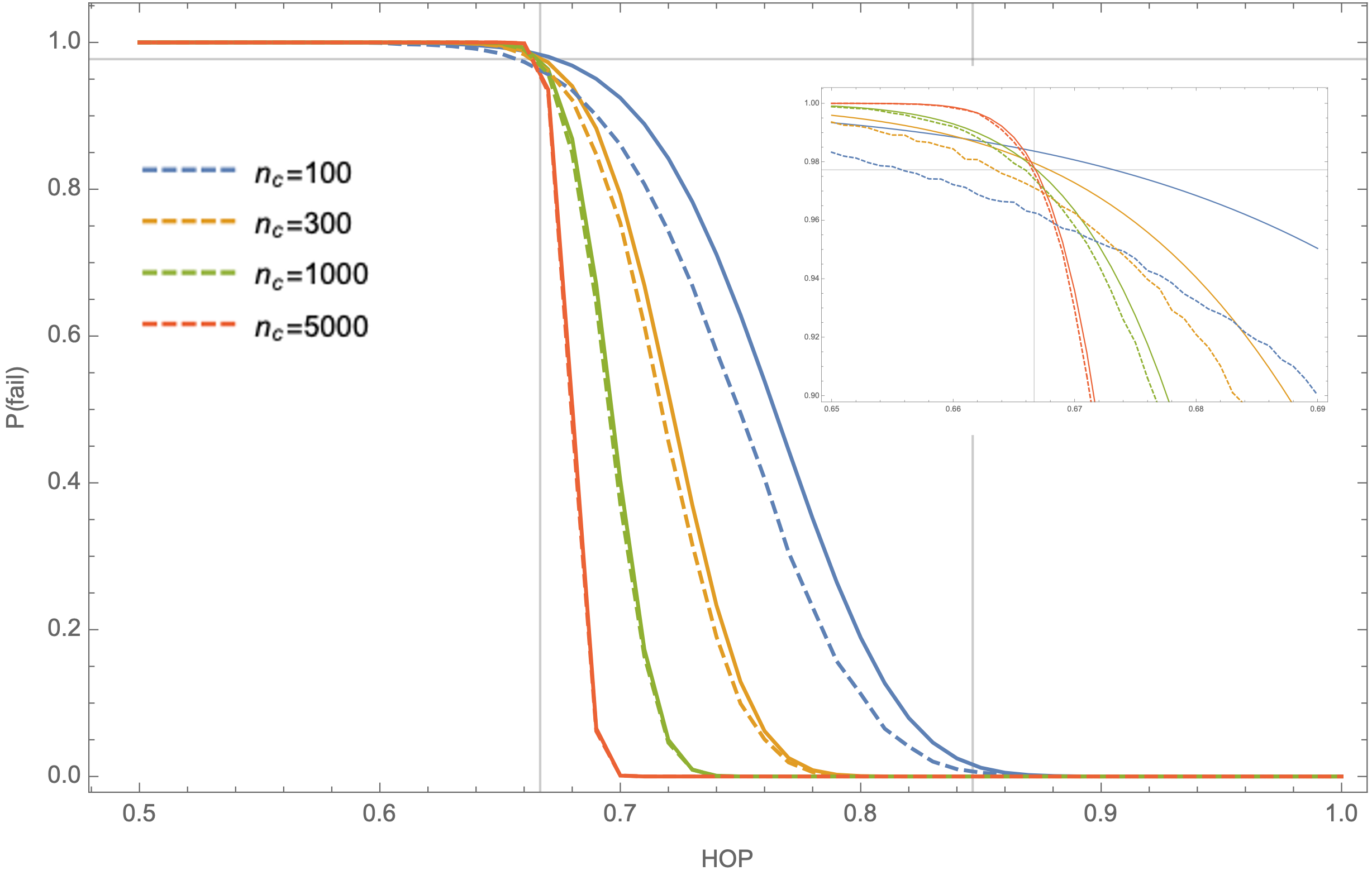}
    \caption{Failure rate as a function of true HOP for the original~\cite{cross2019validating} (solid lines) and bootstrapped~\cite{baldwin:2022} (dashed lines) protocols, for differing number $n_c$ of circuits sampled. Ideally there would be a step-function change at HOP of $2/3$ (denoted by vertical gridline). Inset is a zoom into the important region close to the passing threshold. The original protocol keeps the false-positive rate capped at the scale of the target $\epsilon$ whereas the bootstrap protocol exceeds by a factor $\sim1.6$.}
    \label{fig:negativeratevshop}
\end{figure}

Alternatively, in Ref.~\cite{baldwin:2022} the authors considered a different set of conditions for passing. In this reference, the authors use a bootstrap protocol, which they claim gives a more accurate accounting of the confidence interval false negative. To illustrate this, in Fig.~\ref{fig:negativeratevsnc} and Fig.~\ref{fig:negativeratevshop} we show the probability that the QV benchmark fails when it is intended to fail, i.e., the true HOP in the infinite circuit limit is $\epsilon$ below the $2/3$ threshold, for the two procedures. We see that bootstrapping gives a higher chance of false positives, particularly if fewer circuits are sampled. Based on the clear rules set forth here and in Ref.~\cite{cross2019validating} for the QV benchmark this is \emph{not} the QV benchmark, but should be considered a differently defined benchmark. 

Because of the need to classically simulate the circuit to determine the heavy-outputs, QV is not a scalable benchmarking technique. In \S~\ref{sec:mirror} we show how to overcome this using QV circuits in a mirror benchmarking protocol \cite{proctor2022scalable}.  \\

{\renewcommand{\arraystretch}{1.2}
\begin{table}[h!]
    \centering
    \bigskip
    \begin{tabular}{ |p{4cm}|p{4cm}|} 
    \hline
    \multicolumn{2}{|c|}{\textbf{Rules of quantum volume}} \\
    \hline
    \textbf{Do} & \textbf{Don't} \\ 
    \hline
    Run a minimum of 100 random circuits.
    &
    Use knowledge of ideal output distribution/HOP as cost function to optimize circuit compilation or error suppression.
    \\ 
    \hline
    Run until heavy output probability $h_{U} > 2/3 \pm \sigma $.
    &
    Error mitigation acting on quasi-distribution or expectation values: No measurement mitigation, zero-noise extrapolation, or probabilistic error cancellation.
    \\ 
    \hline
    Run more than a single shot per circuit (at least $1000$ shots per circuit recommended).
    &
    Post-selection based on HOP.
    \\ 
    \hline
    Implement circuit unitary.
    &
    \\ 
    \hline
    \end{tabular}
    \caption{Rules for executing the quantum volume benchmark protocol}
    \label{tab:qv_rules}
    \bigskip
\end{table}}

\subsubsection{CLOPS}
Circuit Layer Operations per Second (CLOPS) \cite{wack2021quality} is an example of a benchmark aimed at measuring the speed of a quantum processor. CLOPS looks at the number of QV layers per second that can be executed from a set of parameterized QV circuits that are updated during run-time without sacrificing the best value obtained for QV. A high CLOPS value indicates that the hardware can perform a large number of operations per second. The expression for CLOPS given in Ref. \cite{wack2021quality} is the following

\begin{equation}
        \text{CLOPS} = \frac{M \times K \times S \times D}{\text{time}\_\text{taken}} 
\end{equation}

\noindent
where $M$ is the number of template circuits, $K$ is the number of parameter updates, $S$ is the number of shots, $D=\log_2 V_Q$ is the number of QV layers and $\text{time}\_\text{taken}$ is the total execution time.

Note that CLOPS is a metric focused on the runtime architecture, circuit execution, and data transfer, mimicking the inner loop of a parametric update flow. As such, it starts the timer after circuits have been compiled to whatever point the platform supports where parameters can still be applied in each iteration. Further processing after this point is captured in the time as well as the time to return the results back to the benchmark code to create the next set of parameters.  Since it does not specify the classical pre-processing necessary to generate the initial circuits or post-processing to generate these parameters, other applications may incur more or less classical overhead than the CLOPS benchmark.\\

\subsubsection{\label{sec:mirror}Mirror circuits and mirror quantum volume}

Mirror circuits \cite{proctor2022measuring} are a general class of quality benchmarks where the circuit of interest is concatenated with its inverse mirror copy (up to a Pauli). Optional gate layers can be inserted at the start, end, and between the two halves of the circuit to adjust its sensitivity to certain types of errors. The advantages of mirror circuits are their efficient classical compilation, simple to quantify outcome, and applicability to circuits of different structures as a benchmark for measuring how well the circuit of interest can be executed on the target device. In particular, mirror randomized benchmarking is an extension of the standard randomized benchmarking protocol and is described in Ref.~\cite{proctor2022scalable}. Because it can be constructed from only single qubit Cliffords and CNOTs while remaining simple to analyze and extract figures of merit from, it is a good scalable benchmark for qubit numbers well beyond what is typically feasible for standard randomized benchmarking. The protocol can also be extended to benchmark universal gate sets \cite{hines2022demonstrating}.

\begin{figure}[]
     \centering
     \begin{subfigure}[]{\columnwidth}
         \centering
         \includegraphics[width=0.8\columnwidth]{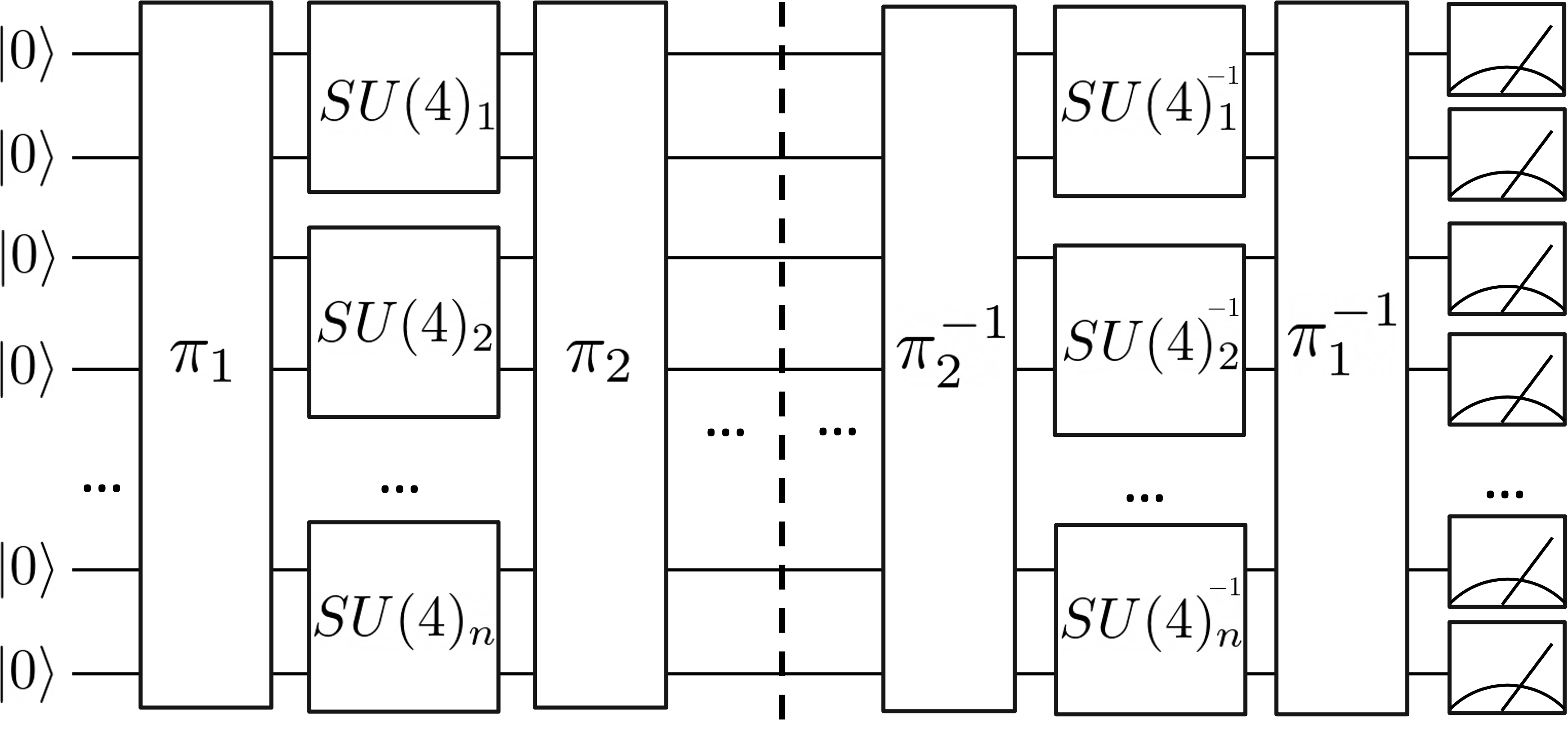}
         \caption{ }
         \label{fig:mqv-circuit}
     \end{subfigure}
     \begin{subfigure}[]{\columnwidth}
         \centering
         \includegraphics[width=0.85\columnwidth]{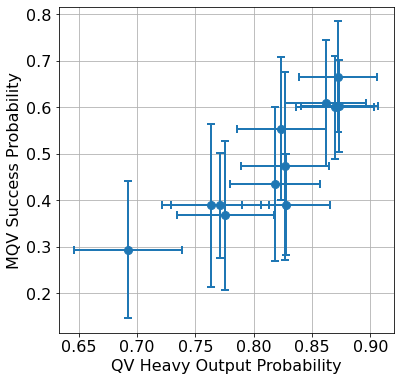}
         \caption{}
         \label{fig:qv-vs-mqv}
     \end{subfigure}
     \hfill
        \caption{(a) The mirror quantum volume circuit, consisting of permutation and $SU(4)$ layers that are inverted about the dashed line of symmetry. (b) A comparison of mirror QV success probabilities versus standard QV HOP over linear 4-qubit strings across the 65-qubit IBM Quantum Ithaca processor. 500 circuits of 1000 shots each were run for each benchmark.}
        \label{fig:mqv}
\end{figure}

While scalability is not an absolute postulate of benchmarks, it is a desirable property.  Mirror circuits are a clear path to scalability, and have been demonstrated as a scalable method for measuring large GHZ states~\cite{wei:2020}. Here we demonstrate a similar method with quantum volume. To mirror quantum volume, we apply the simple transformation in Ref.~\cite{proctor2022measuring} by replacing half of the standard QV circuit with the inverse of the other half, such that the output bit-string is all zeros, as shown in Fig.~\ref{fig:mqv-circuit}. The depth of such a circuit is the same as the depth $d$ of the original QV circuit in the case of even $d$ and is $d+1$ if $d$ is odd. Care should be taken during compilation that the half-circuits do not get reduced down to a circuit of lower depth from simple cancellation. Instead of HOP as the output metric, the mirror QV protocol calculates the success probability of the output bit-string being the same as the ideal bit-string. Depending on the characteristic errors of the target processor, optional Pauli or Clifford twirling layers can also be added to the head or the tail of the entire mirror circuit with minimal classical and quantum overhead.

Mirror QV is significantly faster than standard QV because there is no need to classically simulate the circuits, but the potential downside is that due to its symmetric nature and simple output, mirror QV may be insensitive to errors that do not impact the all-zero output state and systematic errors that would cancel out between the first and inverse halves of the circuit, which are all errors that standard QV may be susceptible to. For the latter errors, one possible scheme for increasing error sensitivity in solid-state and atomic quantum systems is to not use commonplace inverse gate implementations that reverse the sign of the control field, but to instead compile the circuit with the same native physical gate set for both layers. The idea of using sign-inverted gates to cancel coherent error was explored in \cite{zhang2022hidden}, and here we can purposefully \emph{not} use those gates to prevent coherent errors from canceling. But though the basic mirror circuit has these limitations without twirling layers or other optimizations, we find that it's already a good proxy for QV under realistic transmon device errors, as shown in Fig.~\ref{fig:qv-vs-mqv}. Therefore, mirror QV is usable as a fast heuristic for evaluating the potential quality of QV over a large quantum processor.

\subsubsection{\label{sec:application}Application-oriented performance benchmarks}

There have been a number of efforts to collect circuits together that are relevant and/or proxies to expected applications on quantum computers \cite{lubinski2021application, mesman2021qpack, finvzgar2022quark, tomesh2022supermarq, zhang2023characterizing, kurlej2022benchmarking, lubinski2023optimization, kordzanganeh2022benchmarking, mundada2022experimental,li:2022}. Evaluating the output of these circuits is a useful source of system information, however, to rise to the level of benchmark they must include additional information to adhere to our previously discussed guidelines. One such effort was undertaken in Ref.~\cite{lubinski2021application}, which mostly conforms to our rules except that the specific circuits in the benchmark are not clearly prescribed. We use the method of Ref.~\cite{lubinski2021application} later to experimentally demonstrate some of our ideas on benchmarks. Here we give a brief overview of the benchmarking suite in Ref.~\cite{lubinski2021application}. The starting point is a repository of algorithms or subroutines that have been converted into a set of quantum circuits. All the algorithms in the suite start with the qubits with the ground state, the circuit is applied and the qubits are measured. The quality of the circuit is measured by comparing the experimentally obtained sampled-probability distribution $P_{\text{output}}$ with the ideal one $P_{\text{ideal}}$ that can be obtained by simulating the circuit, normalizing with respect to the uniform distribution $P_{\text{uni}}$. This takes the following expression: 

\begin{equation}
        F\left( P_{\text{ideal}}, P_{\text{output}} \right) = \text{max} \left\{ F_{\text{raw}} \left( P_{\text{ideal}}, P_{\text{output}} \right), 0 \right\} 
\end{equation}

\noindent 
with 
\begin{equation}
        F_{\text{raw}}\left( P_{\text{ideal}}, P_{\text{output}} \right) = \frac{F_s \left( P_{\text{ideal}}, P_{\text{output}} \right) - F_s \left( P_{\text{ideal}}, P_{\text{uni}} \right)}{1 - F_s \left( P_{\text{ideal}}, P_{\text{uni}} \right)}
\end{equation}

\noindent
where $F_s \left(P_A, P_B \right) =\left(\sum_x \sqrt{P_A(x) P_B(x)}\right)^2$ is the Hellinger fidelity between two sampled-probability distributions.\\

In Ref.~\cite{lubinski2021application}, it is stated that the benchmark consists of selecting a set of circuits (from the available collection), then selecting a range of qubit numbers to investigate. For a given algorithm  and number of qubits $n$, a set of $C_n$ circuits is generated. $N_{\text{circs}}$ are randomly sampled out of $C_n$. For each of the $N_{\text{circs}}$ circuits, a random sampling of the parameters (if there are any) generates further $N_{\text{params}}$  where any adjustable parameter is fixed by randomly sampling their values. This means that for a certain size $n$, $N_{\text{circs}} \times  N_{\text{params}}$ circuits are generated. Each circuit is transpiled to the native gate-set and run on the hardware. The fidelity of the algorithm at size $n$ is then the average fidelity over these realizations. It is recommended that at least $N_{\text{circs}} = 10$ circuits be chosen. In general, users can run as many algorithms as they like to get a sense of performance on problems of interest with similar structure. For the sake of comparison across devices or iterations of the same device, it is suggested that all algorithms be run.

\section{\label{sec:opt_bench} Optimizing Benchmarks}

\begin{figure}[h]
     \centering
         \includegraphics[width=\columnwidth]{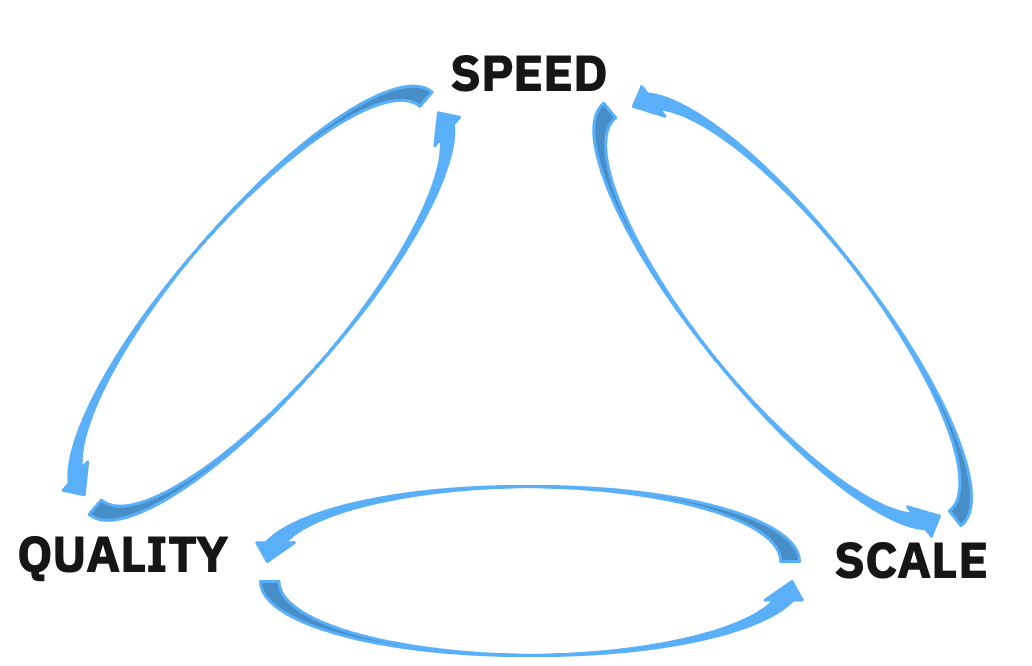}
     \hfill
        \caption{Different attributes of a quantum computer are traded off with each other in error mitigation and correction techniques.}
        \label{fig:sqs_tradeoffs}
\end{figure}

One of our postulates for a benchmark is that it must obey a universal set of rules. Staying within that set of rules, one is free to optimize the benchmarking performance. In fact, there is almost an obligation to do that, so as to avoid misrepresenting the performance of any particular devices, for example by poorly transcribing the gates in a benchmark or by inserting unnecessary idle times. However, one has to be careful when such optimizations start to trade off against the key device attributes we are benchmarking---scale, quality and speed (Fig.~\ref{fig:sqs_tradeoffs}). In particular, error mitigation and correction techniques allow us to trade speed or scale for quality. The most striking example is the trade-off between scale and speed in exchange for quality in error correcting codes \cite{babbush2021focus, chen2022calibrated}, where extra qubits and operations are sacrificed to improve the error rates of so-called logical qubits. Error mitigation techniques also provide a variety of examples where different trade-offs can happen. In probabilistic error cancellation (PEC) \cite{temme2017error, berg2022probabilistic}, an exponential (in the noise strength) number of circuits needs to be run to obtain bias-free estimates of observables, thus trading off speed for quality. In virtual distillation \cite{koczor2021exponential, huggins2021virtual}, powers of the density matrix obtained by evaluating observable on copies of the state give its largest eigenstate, which is assumed to be the noise-free state. Here scale is traded-off with quality. From these examples, it is clear that the term error mitigation has come to capture a plethora of techniques that aim at accounting for errors affecting the results of quantum circuits. In practice, some of these techniques require some amount of classical processing, others require extra quantum circuits to be executed and others need both extra classical processing and quantum circuits. We'll refer to techniques that aim at reducing the occurrence of errors  as error suppression techniques while techniques that correct errors after they happened as error mitigation techniques. Examples of the former are heuristic mapping routines which map a generic quantum circuit to the connectivity of the hardware where it will be executed \cite{li2019tackling}, compilation passes which add dynamic decoupling sequences to idle times of the circuit or the selection of the best performing qubits that support the quantum circuit \cite{nation2022suppressing, murali2019noise}. Although some of these techniques (as the optimal mapper in Ref.~\cite{nannicini2022optimal}) may incur in an exponential overhead in classical computation when the number of qubits is increased, most of them can be run efficiently by limiting their run-time. 

Other techniques may require additional quantum circuits to be run without a noticeable increase in classical computing resources needed to account for errors. One such example is Refs. \cite{van2022model, nation2021scalable}, that incur a constant-overhead in the number of circuit instances needed to run in order to mitigate measurement noise.  However, the precision of the resulting expectation values scales exponentially with readout error rates and the weight of the mitigated observables; they still require exponential quantum resources. 

While others again may require an exponential amount of quantum or classical computation. An example of an exponentially expensive technique in the amount of quantum compute is probabilistic error cancellation \cite{berg2022probabilistic}.

Given these issues, we want to put down a set of rules for optimization. \\

\textbf{Rule 1: Constant time optimizations are allowed and encouraged.} The most prominent of the optimizations falling into this category is that of dynamic decoupling (``error supression''), but this also includes optimal compiling of gates and efficient measurement mitigation. Other optimizations that fall under this scope can be optimal mapping to the device; technically this is exponential in the size of the device, but only needs to be done ones per benchmark run (not per circuit) and there are efficient versions, e.g., not looking at every possible layout. Similarly, optimal pre-compilations (that don't violate Rule 3) and are not per circuit, but costly in time are allowed. \\

\textbf{Rule 2: Mitigation must be reported along with the incurred overhead.} This covers a wide class of techniques (discussed above) that enable an increase in quality by trading off speed or scale. At one extreme, we could imagine even replacing the run on the device by a fully classical simulator. While this would be far outside the spirit of benchmarks (and therefore not allowed, and likely not pass the rules of said benchmark), this would be transparently observed by enforcement of this rule. Furthermore, we need to acknowledge that the results of benchmarking methods that include error mitigation may not be directly comparable to those that do not. Therefore, when comparing results from different devices or different versions of the same device, it's important to take into account the different error mitigation techniques that have been applied. \\

\textbf{Rule 3: Optimizations based on the output of a circuit are forbidden.} While this should be in clear violation of the spirit of benchmarks, here we make it explicit. There can be no optimization of the result or the output based on the knowledge of the what the output of the circuit is expected to be. For example, replacing the circuit with a much simpler version that still obtains the right output and/or post-selecting only the correct outputs. \\

\section{\label{sect:examples} Experimental Examples}

To highlight some of the concepts that we introduced in the previous sections, here we look at experimental examples using the application benchmarking suite and mirror circuits (\S~\ref{sec:application} and \ref{sec:mirror}). 

\subsection{Constant Time Optimizations}

\begin{figure}[]
     \centering
     \begin{subfigure}[t]{\columnwidth}
         \centering
         \includegraphics[width=0.6\columnwidth]{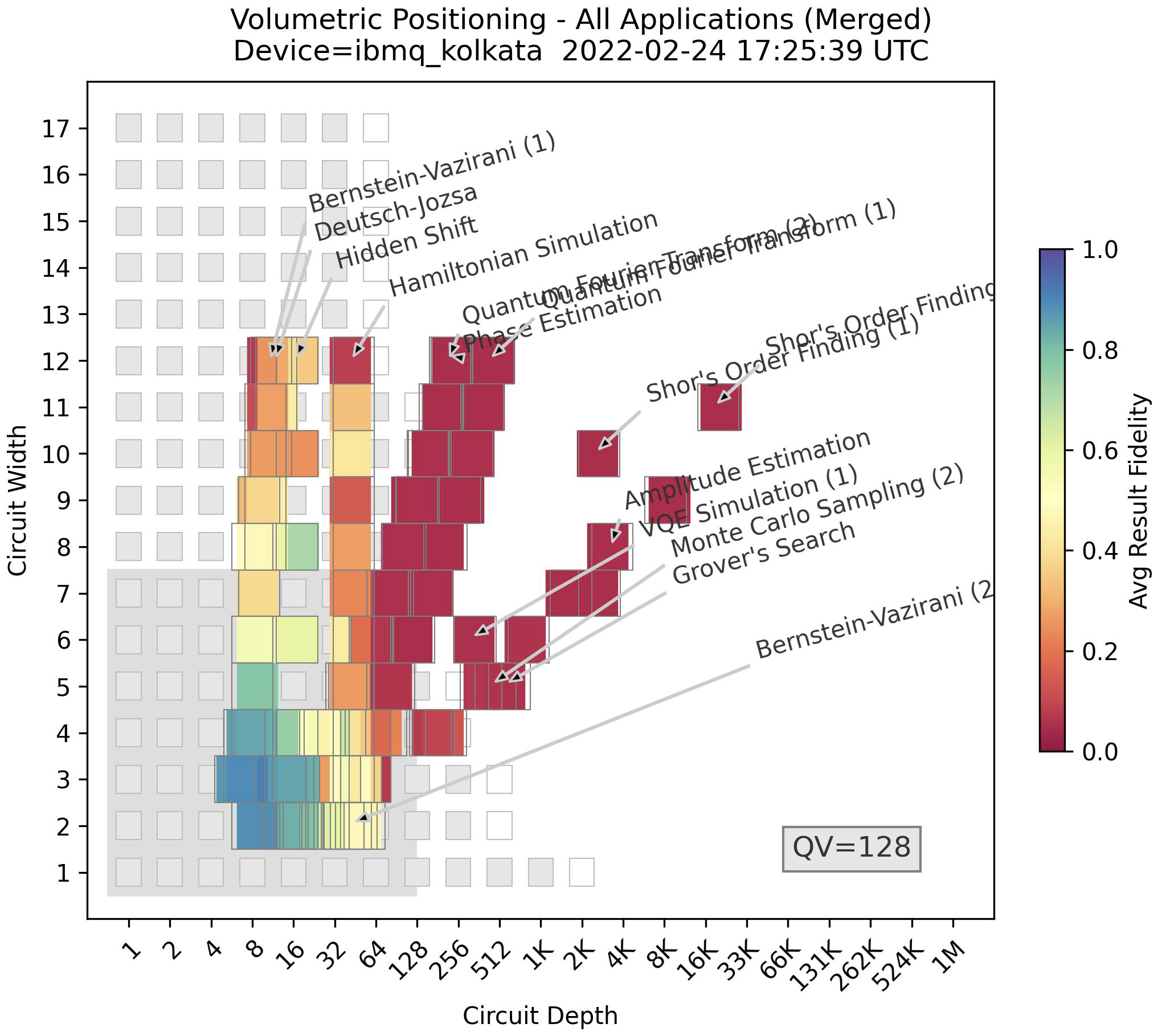}
         \caption{ }
         \label{fig:kolkata-default}
     \end{subfigure}
     \hfill
     \begin{subfigure}[t]{\columnwidth}
         \centering
         \includegraphics[width=0.6\columnwidth]{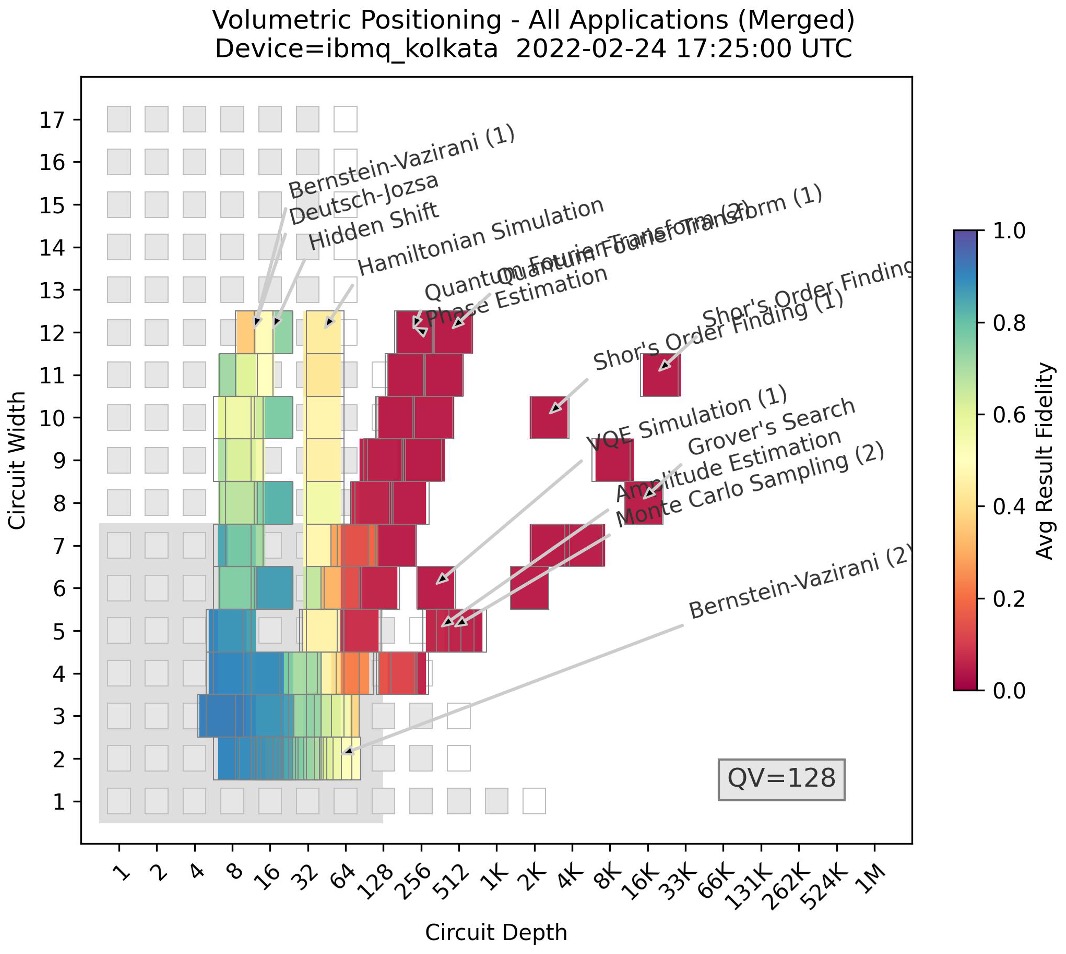}
         \caption{}
         \label{fig:kolkata-layout}
     \end{subfigure}
     
  \medskip

     \begin{subfigure}[t]{\columnwidth}
         \centering
         \includegraphics[width=0.6\columnwidth]{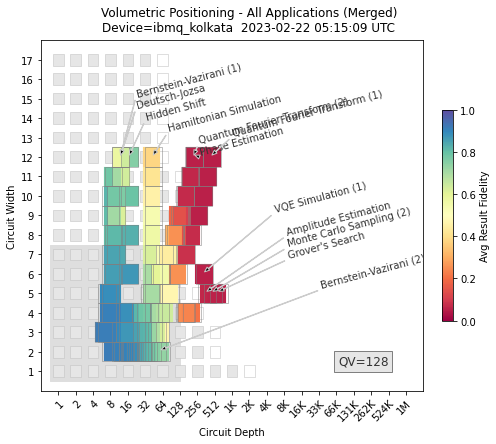}
         \caption{}
         \label{fig:kolkata-layout-dd}
     \end{subfigure}
     \hfill
     \begin{subfigure}[t]{\columnwidth}
         \centering
         \includegraphics[width=0.6\columnwidth]{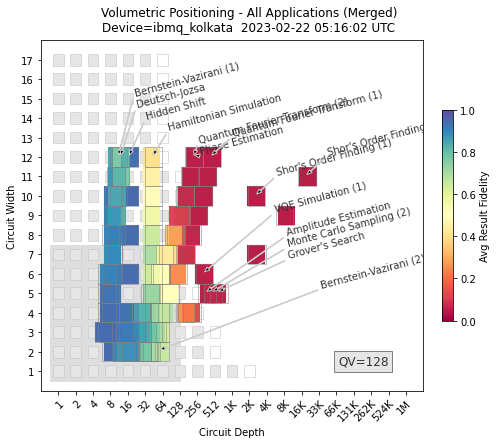}
         \caption{}
         \label{fig:kolkata-layout-dd-mmit}
     \end{subfigure}
        \caption{Comparison of benchmarking results from the QEDC application-oriented suite of Ref. \cite{lubinski2021application} (a) executed with default parameters,  (b) executed by adding layout selection, (c) layout selection and dynamic decoupling and (d) layout selection, dynamic decoupling and measurement error mitigation.}
        \label{fig:qedc_plots}
\end{figure}

It is important to include the cost of different error reduction strategies when determining performance because the increase in quality that they can provide may come at the cost of scale or speed. In Fig.~(\ref{fig:qedc_plots}) we show the results of running the application-oriented benchmark suite of Ref.~(\cite{lubinski2021application}) for constant (or low overhead) time techniques. Each plot shows the compounded change in quality given by sequentially adding these techniques. Over the different plots, a tremendous difference in the quality of certain applications can be observed, illustrating the importance of such optimizations. The circuits are executed on the 27Q IBM Quantum Kolkata device that, at the time of running, had average gate errors of $1.8 \times 10^{-4}$ and  $6.8 \times 10^{-3}$ for single-qubit and two-qubit gates, respectively, as  measured  by the randomized benchmarking protocol. In addition, $\rm T_{1} \sim 134 \mu s$, $\rm T_{2} \sim 92 \mu s$ and $\sim 1\%$ readout assignment error.

 More specifically, the results obtained in Fig.~(\ref{fig:kolkata-default}) is what can be achieved with default execution settings. This only includes attempts to reduce the number of SWAPs used for routing, the mapping of a quantum circuit with all-to-all connectivity to the topology of the device. In Fig.~(\ref{fig:kolkata-layout}), we introduce an error suppression techniques which looks for the best layout selection. This takes the quantum circuit which we are interested in running and the device calibration data to pick the best qubits to run the circuit without altering (no extra routing involved) the circuit itself. The routine is based on isomorphic sub-graph methods which add little overhead relative to the compilation time. An in-depth study of the layout mapper used here was done in Ref. \cite{nation2022suppressing}. Figure~(\ref{fig:kolkata-layout-dd}) shows that further improvements that can be obtained when a simple (X-X) dynamic decoupling sequence is added to the idle times of the circuit \cite{pokharel2018demonstration}. This also involves simple graph traversing routines and requires little classical overhead to be implemented. Finally, we add the measurement error mitigation technique of Ref.~(\cite{nation2021scalable}) to the pipeline and obtained the results of Fig.~(\ref{fig:kolkata-layout-dd-mmit}).

\begin{figure}[]
     \centering
         \centering
         \includegraphics[width=\columnwidth]{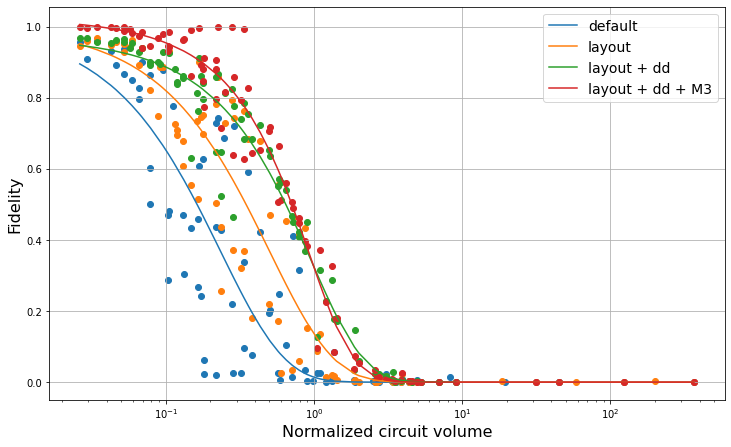}
        \caption{Alternative visualization of the data in Fig. \ref{fig:qedc_plots}, where the fidelity of each application is plotted as a function of the circuits space-time volume normalized to the average volume of a QV128 circuit on IBM Quantum Kolkata device. Different colors correspond to different set of error mitigation techniques used. Solid lines are a logistic curve fit to the data points.}
        \label{fig:kolkata-norm-circ-vol}

\end{figure}

Figure~(\ref{fig:kolkata-norm-circ-vol}) offers a different way to look at the information contained in Fig.~(\ref{fig:qedc_plots}). On the vertical axis are the polarization fidelities (defined in Ref. \cite{lubinski2021application}) for each of the applications while on the horizontal axis is their normalized circuit volume. The normalized circuit volume is defined as the ratio between the space-time volume of the application circuit ($\text{width} \times \text{depth}$) to the volume of an average quantum volume circuit corresponding to the reported quantum volume of the device. A logistic curve $y=\frac{a}{1+b c^{-dx}}$ is fitted to the data points (with $a, \,b, \, c, \,d$ parameters of the fit) showing initial exponential decay in fidelity with the size of the circuit. Different curves correspond to the different plots in Fig.~(\ref{fig:qedc_plots}), where a different set of error mitigation techniques applied. Interestingly, the fidelity of the best error suppressed results decay to ~$1/e$ for circuits that have a volume equal to that of the average quantum volume circuit. This could suggest a way to determine the success of an application based on its area, assuming some amount of error suppression is used. However, the inaccuracy of the fit leads us to shy away from taking it as a metric for the benchmark.  

\begin{figure}[h!]
     \centering
     \begin{subfigure}[t]{0.49\columnwidth}
         \centering
         \includegraphics[height=\linewidth]{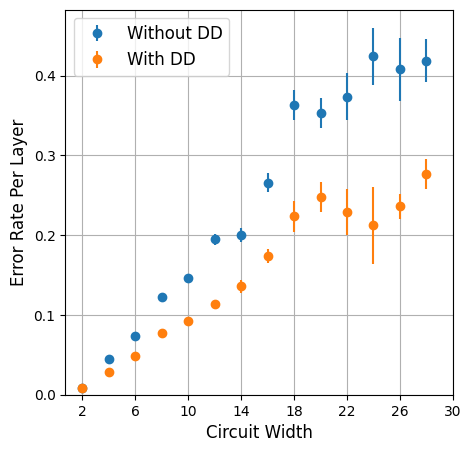}
         \caption{ }
         \label{fig:mrb-epl}
     \end{subfigure}
     \begin{subfigure}[t]{0.49\columnwidth}
         \centering
         \includegraphics[height=\linewidth]{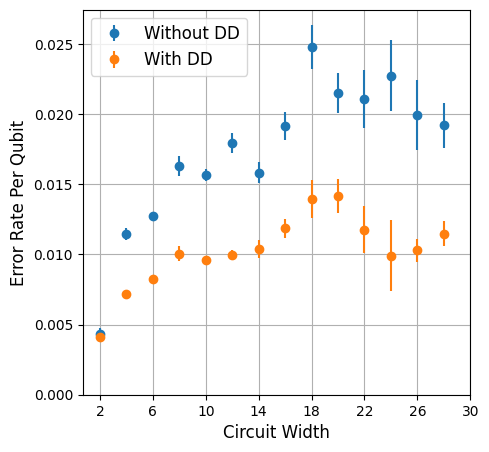}
         \caption{}
         \label{fig:mrb-epq}
     \end{subfigure}
        \caption{The (a) average layer error rate and (b) average qubit error rate (calculated using the method in Ref.~\cite{hines2022demonstrating}) for mirror randomized benchmarking with and without dynamic decoupling (DD) run on linear strings up to 28 qubits on IBM Quantum Ithaca device. The universal gate set chosen was Haar-random $\mathbb{SU}(2)$ single qubits gates and CXs, with two-qubit gate density $\sfrac{1}{4}$.}
        \label{fig:mrb-data}
\end{figure}

Another example of how error suppression techniques can improve the quality measured by a benchmark is shown in Fig.~(\ref{fig:mrb-data}), where we run mirror randomized benchmarking with a universal gate set on the IBM Quantum Ithaca device, both with and without dynamic decoupling. Reference~(\cite{hines2022demonstrating}) had concluded from similar data, taken with the same gate set and two-qubit gate density but without dynamic decoupling, that the observed increase in errors as circuit width increased was due to the rise in cross-talk. However, here we show that the application of simple X-X dynamic decoupling on idle qubits greatly mitigated these errors. Indeed, dynamic decoupling is particularly effective on sparsely structured circuits where many qubits idle for long durations, which is especially the case for mirror randomized benchmarking circuits run with low two-qubit gate density.

\subsection{\label{subsec:error_mitigation} Error Mitigation with exponential overhead}
\begin{figure}[h!]
     \centering
     \begin{subfigure}[t]{\columnwidth}
        \centering
         \includegraphics[width=0.8\columnwidth]{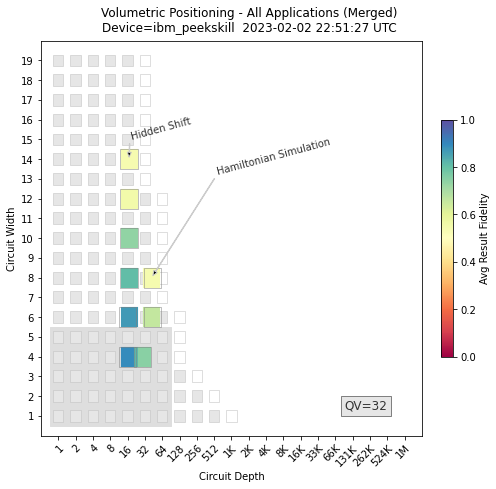}
         \caption{ }
         \label{fig:peekskill-no-pec}
     \end{subfigure}
     \hfill
     \begin{subfigure}[t]{\columnwidth}
        \centering
         \includegraphics[width=0.8\columnwidth]{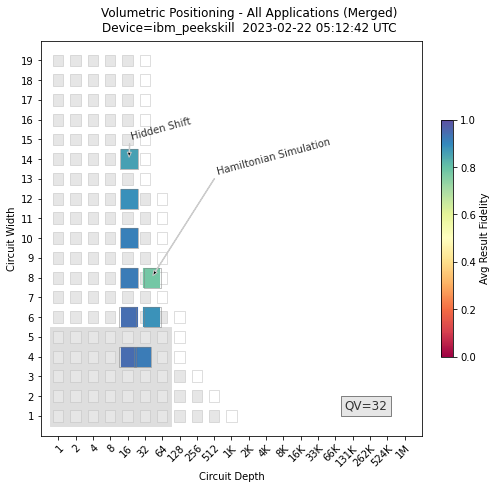}
         \caption{}
         \label{fig:peekskill-pec}
     \end{subfigure}
        \caption{Results for the Hidden-shift and Hamiltonian simulation algorithm from the QEDC application-oriented suite (a) executed with error suppression techniques and (b) executed by mitigating with probabilistic error cancellation.}
        \label{fig:pec_plots}
\end{figure}

When considering the use of error mitigation techniques for improving benchmark quality, it is of utmost importance to clearly identify the trade-offs implied. We give an example of this in Fig.~(\ref{fig:pec_plots}) where we show a comparison between the quality of two algorithms of the collection in Ref.~(\cite{lubinski2021application}). In Fig.~(\ref{fig:peekskill-no-pec}), the same set of error suppression and mitigation techniques used in Fig.~(\ref{fig:kolkata-layout-dd-mmit}) is applied. In Fig.~(\ref{fig:peekskill-pec}) PEC was used on the IBM Quantum Peekskill system through the Qiskit Runtime primitives \cite{IBM}. At the time of the experiment, device calibrations reported the following averaged data: $\bar{T}_1 = 364 \mu s$, $\bar{T}_2 = 349 \mu s$, $1.2\%$ readout assignment error, $1.5\times 10^{-4}$ and $5.3 \times 10^{-3}$ single and two-qubit gate infidelity, respectively, as measured by the randomized benchmarking protocol. In the case of the results obtained with PEC, Near perfect fidelity was achieved for the algorithms tested. In particular, we focused on the ones which are most amenable for this error mitigation technique, which means that their circuit is structured in such a way that a repetitive pattern of simultaneous two-qubit gates can be identified. This makes the technique more effective, given that learning must be done on a timescale over which device noise is approximately static. Because of the exponential amount of resources in the noise strength of the circuit needed for running PEC, it is obvious that the plot in Fig.~(\ref{fig:peekskill-pec}) can be deceiving as only the final quality of the results obtained is shown, without reference to the time it took. In fact, each data point on the plot required running a noise learning procedure taking 5040 circuits at 1024 shots for each unique layer of two-qubit gates present in the circuit. On top of the learning overhead, the mitigated expectation value is calculated from a collection of 12000 circuits at 1024 shots. Compared to the unmitigated case, where only one circuit with 1024 shots was run, this means we paid a $\sim22000\times$-$32000\times$ cost in speed for executing the benchmarks with error mitigation.

\subsection{\label{subsec:gaming_strategies} Using the Circuit Output}

For certain problems, like in the quantum chemistry setting, one of the available error mitigation technique is to make use of information that is available a-priori, such as the symmetry conservation properties in a particular problem, to post-select noisy results and discard ones that are in disagreement with what we know. In these settings, it seems reasonable to make use of known information to reduce errors in the final results. However, when considering the situation of measuring the performance of a quantum device, this could give misleading results.

One example is the polarization fidelity defined in Ref.~(\cite{lubinski2021application}). The measure is constructed in such a way that the expected result is a probability distribution on bit-strings that is peaked on a single bit-string. As a consequence, every application in the suite is built so that the output will be unique. In a spirit similar to the one described above, we could embrace the principle of considering low-frequency bit-strings as results produced by noise and thus decide to drop them from the final results. In Table \ref{tab:bv}, we give an example of how the polarization fidelity is affected by dropping low-frequency bit-strings. By sacrificing a moderate amount of shots,  we can improve the polarization fidelity up to a perfect value, giving the false impression of a well-performing hardware. 

{\renewcommand{\arraystretch}{1.2}
\begin{table}[h!]
    \centering
    \bigskip
    \begin{tabular}{ |p{6cm}|p{2cm}|} 
    \hline
    \multicolumn{2}{|c|}{\textbf{Bernstein-Vazirani algorithm}} \\
    \hline
    \textbf{Counts} & \textbf{Polarization Fidelity} \\ 
    \hline
    \texttt{`0111'}: 3,  \texttt{`1010'}: 3,  \texttt{`0110'}: 818, \texttt{`0100'}: 16, \texttt{`1100'}: 1, \texttt{`1110'}: 13, \texttt{`0000'}: 5, \texttt{`0010'}: 141 &  0.81 \\ 
    \hline
    \texttt{`0010'}: 141 \texttt{`0110'}: 818 & 0.84  \\
    \hline
    \texttt{`0110'}: 818 & 1.0  \\ 
    \hline
    \end{tabular}
    \caption{Filtering low-frequency outcomes can artificially improve the polarization fidelity.}
    \label{tab:bv}
    \bigskip
\end{table}}

\section{\label{sec:conclusions}Conclusions}
We have given a definition of what are the characteristics that define quantum benchmarks. In particular, we highlighted the fact that a benchmark should have a randomized component, with the final result averaged over realizations. It should have well-defined rules for executing the protocol to leave no ambiguity on the procedure. As many of the key attributes of the device should be captured, giving a holistic picture of performance. Finally, the protocol should be platform-independent. An example of how a change in the rules of a benchmark can lead to different benchmark results is also given for the case of QV.

A distinction between benchmarking and diagnostic protocols was discussed, with the goal of disentangling the different aspects that are emphasized by the different methods. Benchmarking protocols measure average performance of a device while diagnostics measure success in specific contexts. Benchmarks inform on average performances and provide results that can be compared across devices and generations. Diagnostic methods give insight on the performance of a device on similarly structured problems and may benefit from the use of mitigation techniques.

Finally, we showed how the quality in diagnostic methods can be improved with the use of various error suppression and mitigation techniques. Efficient constant-time optimization techniques can greatly enhance the quality measured in the benchmarks with a minor overhead. While error mitigation techniques with an exponential overhead can improve, almost arbitrarily, the results. It is thus of utmost important to report which optimization techniques were used to allow for reproducible results. Furthermore, we explore the existence of gaming strategies, like low-frequency outcome filtering, that can artificially improve the quality of results.

\begin{acknowledgments}
The authors would like to thank Johannes Greiner, Luke Govia, Alireza Seif and Doug McClure for insightful discussions, and Francisco Rilloraza and Albert Zhu for developing the Qiskit Experiments-based software to run mirror circuit experiments. HZ and DCM acknowledge support from the Army Research Office under QCISS (W911NF-21-1-0002). The views and conclusions contained in this document are those of the authors and should not be interpreted as representing the official policies, either expressed or implied, of the Army Research Office or the U.S. Government. The U.S. Government is authorized to reproduce and distribute reprints for Government purposes notwithstanding any copyright notation herein.
\end{acknowledgments}

\bibliography{main}

\end{document}